\documentclass{ws-procs975x65}            % default, citations in superscript

\begin{document}
\title{Lepton Number Violation and the Baryon Asymmetry of the Universe }

\author{Julia Harz, Wei-Chih Huang}

\address{
Department of Physics and Astronomy, University College London,\\
London WC1E 6BT, United Kingdom}

\author{Heinrich P\"as}

\address{
Fakult\"at f\"ur Physik, Technische Universit\"at Dortmund,\\ D-44221 Dortmund, Germany}
%$^*$E-mail: heinrich.paes@uni-dortmund.de

%\author{Julia Harz}        \email{j.harz@ucl.ac.uk}\affiliation{\AddrUCL}
%\author{Wei-Chih Huang}    \email{wei-chih.huang@ucl.ac.uk}\affiliation{\AddrUCL}

\begin{abstract}
Neutrinoless double beta decay, lepton number violating collider processes and the Baryon Asymmetry
of the Universe (BAU) are intimately related. In particular
lepton number violating processes at low energies in combination with sphaleron transitions will
typically erase any pre-existing baryon asymmetry of the Universe. In this contribution we briefly review
the tight connection between neutrinoless double beta decay, lepton number violating processes at the
LHC and constraints from successful baryogenesis. We argue that far-reaching conclusions can be drawn unless
the baryon asymmetry is stabilized via some newly introduced mechanism.
\end{abstract}


\bodymatter

\newpage

\section{Introduction}
The discovery of neutrino masses is typically understood as a hint for
physics beyond the Standard Model (SM). Intimately related to this link
is the question whether lepton number is conserved or broken.
After all, neutrino masses can be realized in two different ways, either
as Majorana masses $\overline{\nu_L^C}\nu_L$ or as Dirac masses $\overline{\nu_L}\nu_R +
\overline{\nu_R}\nu_L$. In the first case, lepton number is broken. In the
latter case the newly introduced 
right-handed neutrino is an SM singlet so that a  
Majorana mass $\overline{\nu_R^C} \nu_R$ is allowed by the SM symmetry. So either
lepton number is broken again or this operator has to be forbidden by a new 
symmetry.
In this sense the problem how neutrino masses are related to physics beyond
the Standard Model boils down to the question whether the accidental lepton 
number conservation in the Standard Model is enforced by a new symmetry or
violated by LNV operators. 

\begin{figure}[!ht]
\centering
\includegraphics[width=3.5in]{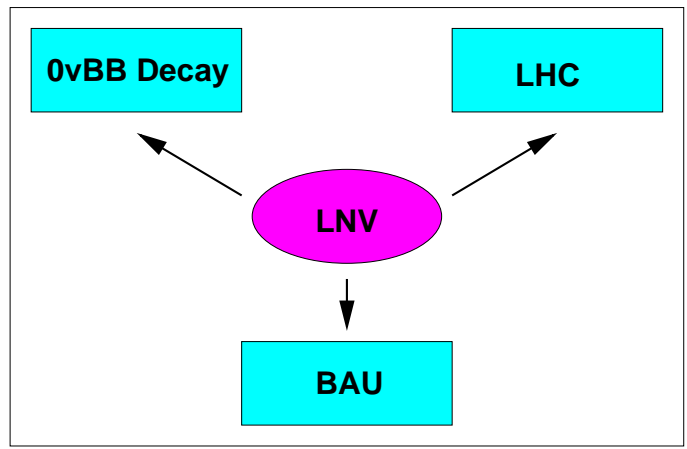}
\caption{Neutrinoless double beta decay, lepton number violating processes at the LHC and the generation and survival of the
Baryon Asymmetry of the Universe are closely interrelated.}
\label{LNV}
\end{figure}

LNV can be searched for directly for example in neutrinoless double beta decay
or at colliders. Moreover, lepton number violating interactions can be important
in cosmology where they can both wash out or create the baryon asymmetry of the
Universe (BAU). These apparently different phenomena are thus closely related
(see Fig.~\ref{LNV}) as will be discussed in the following.

\section{Probing Lepton Number Violation with Neutrinoless Double Beta Decay}

A sensitive probe
of low energy lepton number violation is neutrinoless double beta decay ($0\nu\beta\beta$),
the simultaneous transition of two neutrons into two protons and two electrons, without emission of any anti-neutrinos:
\begin{equation}
2n \rightarrow 2 p + 2e^-.
\end{equation}
While the most prominent decay mode is triggered by a massive Majorana neutrino being exchanged between Standard Model (SM)
$V-A$ vertices, providing a bound on the effective Majorana neutrinos mass
\begin{equation}
\label{eq:defmeff}
\langle m_{\nu}\rangle = \sum_j U_{ej}^2 m_j \equiv m_{ee},
\end{equation}
in the sub-eV range, 
in principle any operator violating lepton number by two units and transforming two neutrons into
two protons, two electrons and nothing else will induce the decay.

\begin{figure}[!ht]
\centering
\includegraphics[width=4in]{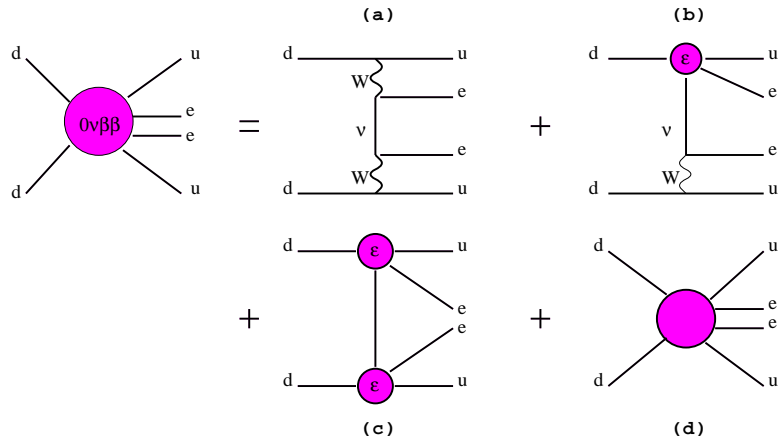}
\caption{Mechanisms for neutrinoless double beta decay:  the most general effective operator triggering the decay can be decomposed
into diagrams with SM vertices and effective vertices being point-like at the nuclear Fermi scale.
(From \protect{\cite{Pas:1999fc}}.)}
\label{general}
\end{figure}

As discussed in detail in~\cite{Pas:1999fc,Pas:2000vn}, the most general operator triggering the decay can be parametrized
in terms of effective couplings $\epsilon$ as shown in Fig.~\ref{general}. 
The diagram depicts the exchange of a light Majorana neutrino between two SM vertices
(contribution a), the exchange of
a light Majorana neutrino between an SM vertex and an effective operator which is pointlike at the nuclear Fermi momentum scale
${\cal O}(100~{\rm MeV})$ (contribution b) and a short-range contribution triggered by a single dimension 9 operator being pointlike at
the Fermi momentum scale (contribution d). Contribution c) which contains two non-SM vertices can be neglected when compared to contribution b). 
The most general decay rate contains all combinations of leptonic and hadronic currents induced by the operators
\begin{equation}
{\cal O}_{V\mp A} = \gamma^{\mu}(1\mp \gamma_5), 
{\cal O}_{S \mp P} = (1 \mp \gamma_5),        
{\cal O}_{T_{L/R}} = \frac{i}{2}[\gamma_{\mu},\gamma_{\nu}](1\mp \gamma_5), 
\label{ops}
\end{equation}
allowed by Lorentz invariance.

Examples for contribution b) are the Leptoquark and SUSY accompanied decay modes,
examples for contribution d) are decay modes where only SUSY particles or heavy neutrinos and gauge bosons
in left-right-symmetric models are exchanged between the decaying nucleons, for a recent overview see~\cite{Deppisch:2012nb}. 
Present experiments have a sensitivity to the effective couplings of 
\begin{equation}
\epsilon < ({\rm few}) \times (10^{-7}-10^{-10}).
\end{equation}
For the $d=9$ operator triggering the contribution d) it can be estimated that an observation of $0\nu\beta\beta$ decay with present-day
experiments would involve TeV scale particles and thus would offer good chances to see new physics associated with LNV at the
LHC. A crucial prerequisite for such a conclusion is of course a possibility to discriminate among the various mechanisms which may
be responsible for the decay. This is a difficult task but may be possible at least for some of the mechanisms by 
observing neutrinoless double beta decay in multiple isotopes \cite{Deppisch:2006hb,Gehman:2007qg}  
or by measuring the decay distribution, for example in the SuperNEMO
experiment \cite{Arnold:2010tu}. Another possibility to discriminate between various short range contributions to neutrinoless
double beta decay at the LHC itself is to identify the invariant mass peaks of particles produced resonantly in the intermediate state or
to analyze the charge asymmetry between final states involving particles and/or anti-particles
\cite{Helo:2013dla,Helo:2013ika}.

%-------------------------------------------------------------------------
\section{Neutrinoless Double Beta Decay at the LHC}
%-------------------------------------------------------------------------

\begin{figure}[!ht]
\centering
\includegraphics[width=3in]{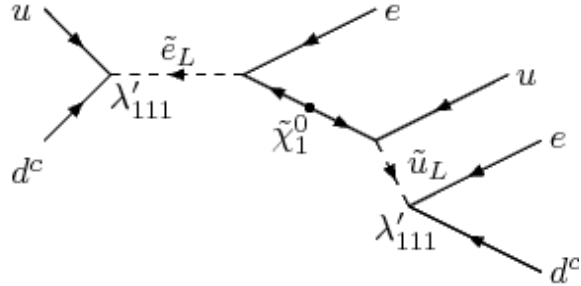}
\caption{Neutrinoless double beta decay at the LHC: the case for R-Parity violation.  Two quarks in the initial state are
converted into a same-sign di-lepton signal and two jets.
(From \protect{\cite{Allanach:2009iv}}.)}
\label{rpvlhc}
\end{figure}

While neutrinoless double beta decay is the prime probe for massive Majorana neutrinos, lepton number violation in general can be searched for
also in collider processes. Indeed, as
has been discussed for example for the special cases of left-right symmetric models \cite{Keung:1983uu,Tello:2010am} and $R$-parity violating supersymmetry \cite{Allanach:2009iv,Allanach:2009xx} the short range contribution d) can easily be crossed into a diagram with two quarks in the initial state
where resonant production of a heavy particle leads to a same-sign dilepton signature plus two jets at the LHC, see Fig.~\ref{rpvlhc}. 
In order to discuss the LHC bounds in a model-independent way similar to the effective field theory approach of \cite{Pas:1999fc,Pas:2000vn},
it is necessary to specify, which particles are produced in the process which requires a decomposition of the d=9 operator. Such a 
decomposition has been worked out in \cite{Bonnet:2012kh} where two different topologies (topology 1 
with two fermions and a boson in the internal lines and topology 2 with an internal 3-boson-vertex) have been specified.
This decomposition was applied to the LHC analogue of $0\nu\beta\beta$ decay and first results for topology 1 
have been derived in  \cite{Helo:2013dla,Helo:2013ika}. The conclusion reached was that with the exception of leptoquark exchange, the 
LHC was typically more sensitive than $0\nu\beta\beta$ decay on the short range operators.  Thus one could infer that typically and with some exceptions

\begin{itemize}

\item
{\it Either}
an observation of  $0\nu\beta\beta$ decay would imply an LHC signal of LNV as well. In turn, no sign of LNV at the LHC would
exclude an  observation of  $0\nu\beta\beta$ decay.

\item
{\it Or} $0\nu\beta\beta$ decay would be triggered by a long-range mechanism a) or b).

\end{itemize}

%----------------------------------------------------
\section{Baryon Asymmetry Washout}
%----------------------------------------------------

An observation of lepton number violation at low energies has important consequences for a pre-existing lepton asymmetry in the
Universe. For example, the prominent leptogenesis scenario for a generation of the baryon asymmetry of the Universe
assumes a lepton number (or $B-L$) and CP asymmetry created in the decays of heavy Majorana neutrinos in the early Universe, which later on
is converted into a baryon asymmetry by  the non-perturbative $B+L$ violating sphaleron transitions
present in the SM. Obviously, such a lepton asymmetry can be washed out by lepton number violating interactions, and  
indeed in \cite{Deppisch:2013jxa} it has been pointed out  that any observation of lepton number violation at the LHC will falsify high-scale leptogenesis. 

The basic argument is that the observation of LNV at the LHC 
will yield a lower bound on the washout factor for the lepton asymmetry in the early Universe. It is easy to see that
this argument can be
extended even further: 

 Just like the combination of $B-L$ violating $\nu_R$ decays in leptogenesis with $B+L$ violating sphaleron  processes can produce a baryon asymmetry, $B-L$ violation observed e.g. at the LHC or elsewhere in combination with $B+L$ violating sphaleron  processes
will lead to a washout of any pre-existing baryon asymmetry, irrespective of the concrete mechanism of baryogenesis.

Combining this argument with the results of~\cite{Helo:2013dla,Helo:2013ika} discussed above,  
one can argue that an observation of short-range $0\nu\beta\beta$ decay will typically imply that LNV processes should be detected at the LHC as well, and this in turn will falsify leptogenesis and in general any high-scale scenario of baryogenesis. 

Indeed, such arguments are not new. They have been first discussed in 
\cite{Fukugita:1990gb} and later on used e.g. to constrain 
neutrino Majorana masses \cite{Gelmini:1992bz},
light lepton number violating sneutrinos
\cite{KlapdorKleingrothaus:1999bd} or Majorana mass terms for 4th generation neutrino states \cite{Hollenberg:2011kq}.
However, only quite recently it has been realized 
in \cite{Deppisch:2015yqa} that the argument can be shown to apply for all short range contributions d) and also for the long-range contribution b) in Fig.~\ref{general}. 
It has been shown that
the $\Delta L = 2$ processes induced by the operator $\mathcal{O}_D$ can be considered to be in equilibrium and the washout of the lepton asymmetry is effective if 
\begin{equation}
  \frac{\Gamma_W}{H} 
	= c_D' \frac{\Lambda_\text{Pl}}{\Lambda_D}\left(\frac{T}{\Lambda_D}\right)^{2D-9} \gtrsim 1,
\end{equation}
where $\Lambda_D$ is the scale of the associated effective operator (assumed to be generated at tree level)
from Eq.~\ref{ops} and 
$c_D'$ being a prefactor of order  ${\cal O}(10^{-3}-1)$.

Thus the far-reaching and strong
conclusion can be drawn that the observation of {\it any} new physics mechanism (i.e. not the mass mechanism)
of neutrinoless double beta decay will typically exclude {\it any} high-scale generation of the baryon asymmetry of the Universe.

Even more recently further studies have been published which analyze the
relation of lepton number violation and the Baryon Asymmetry of the Universe in concrete models such as left-right symmetry or low energy seesaw models \cite{Dhuria:2015cfa,Dev:2015vra}. 

%%%%%%%%%%%%%%
\section{Loopholes}
%%%%%%%%%%%%%%

\begin{figure}[t!]
\centering
\includegraphics[clip,width=0.6\linewidth]{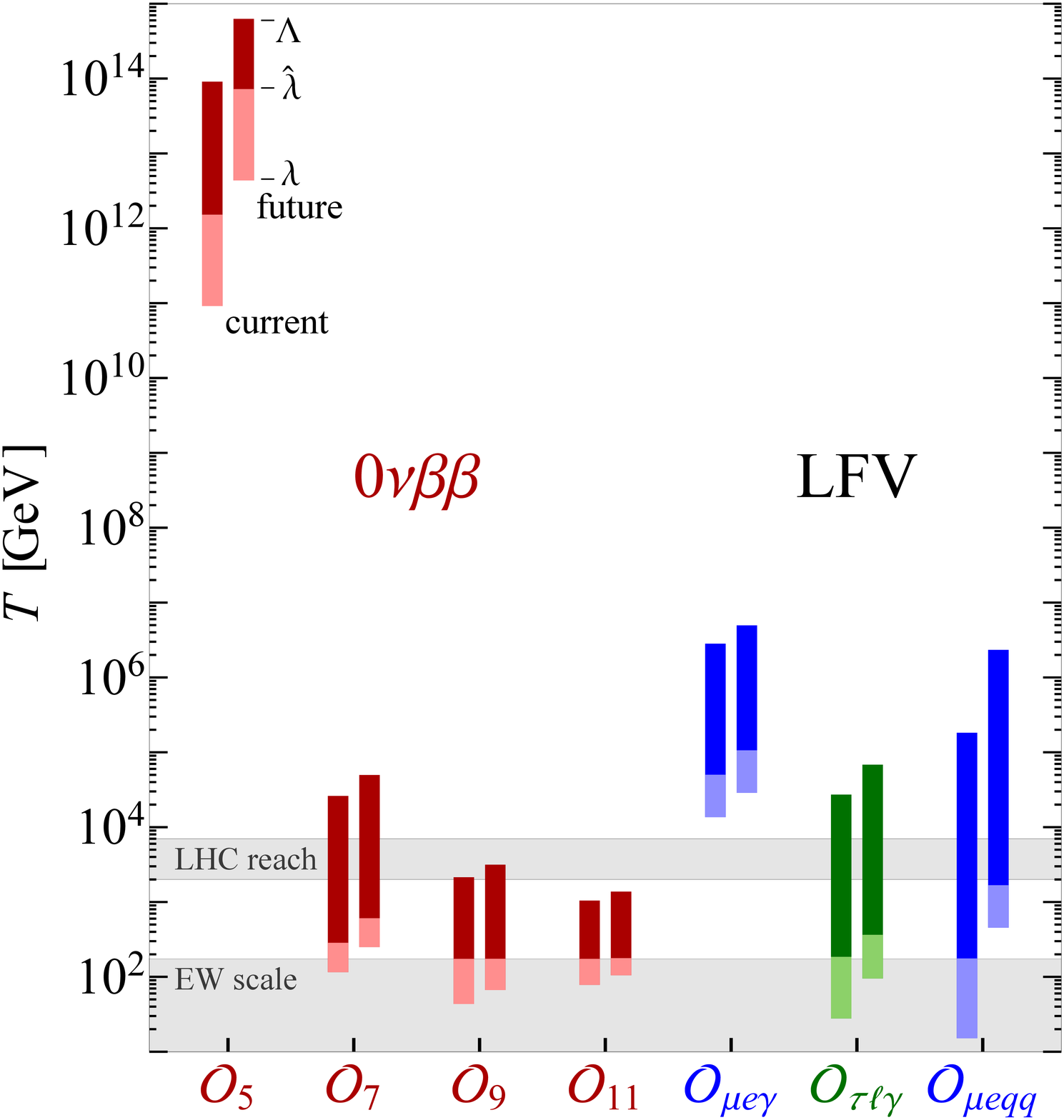}
\caption{Temperature intervals where the given LNV and LFV operators are in equilibrium, assumed that the corresponding process is observed at the current (future) experimental sensitivity.
(From \protect{\cite{Deppisch:2015yqa}}.)}
\label{fig:ranges}  
\end{figure}

Of course these arguments are rather general and various loopholes exist in specific models. 
These include:

\begin{itemize}

\item
Scenarios where LNV is confined to a specific flavor sector only. For example, $0\nu\beta\beta$ decay probes $\Delta L_e=2$ LNV, only.
It may be possible for example that lepton number could still be conserved in the $\tau$ flavor which is not necessarily in
equilibrium with the $e$ and $\mu$ flavors in the early Universe \cite{Deppisch:2013jxa}.
It has been discussed in \cite{Deppisch:2015yqa}, however, that an observation of LFV decays such as $\tau \rightarrow \mu \gamma$
may require LFV couplings large enough to wash out such a flavor specific lepton asymmetry when combined with LNV observed in
a different flavor sector. In Fig.~\ref{fig:ranges}, the
temperature intervals are shown where two individual flavor number asymmetries are equilibrated by LFV processes. When this interval overlaps with the 
$\Delta L = 2$ washout interval of one net flavor number (i.e. electron number if $0\nu\beta\beta$ is observed), the net number of the other flavor will be efficiently washed out as well. As can be seen,  if $\tau\to \ell\gamma$ or $\mu-e$ conversion in nuclei was observed, the involved flavors would be equilibrated around the same temperatures as the washout from the LNV operators 

\item
Models with hidden sectors, new symmetries and/or conserved charges may stabilize a baryon asymmetry against LNV washout as suggested for the example of
hypercharge in \cite{Antaramian:1993nt}.

\item
Lepton number may be broken at a scale below the electroweak phase transition where sphalerons are no longer active.

\end{itemize}

It should be realized though that in general an observation of low energy LNV would invalidate any high-scale generation
of the baryon asymmetry
and that 
the aforementioned protection mechanisms should be addressed explicitly in any model combining low-scale LNV with high-scale baryogenesis.

%

%%%%%%%%%%%%%%
\section{Conclusions}
%%%%%%%%%%%%%%

\begin{figure}[!ht]
\centering
\includegraphics[width=5in]{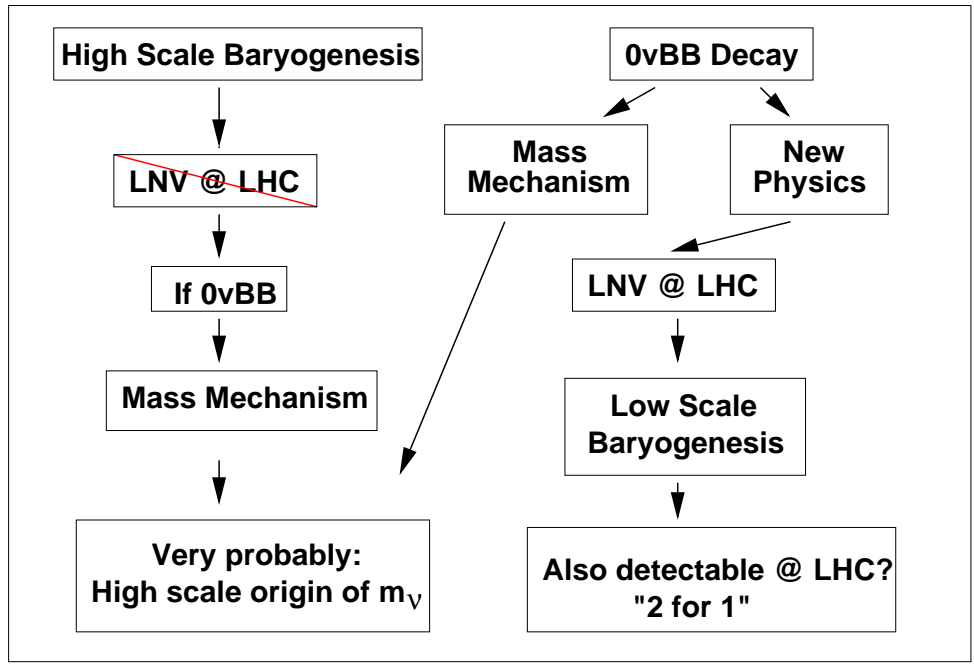}
\caption{Conclusions as a logic tree: a discovery of lepton number violation at low energies will have far-reaching consequences
for the origin of the Baryon Asymmetry of the Universe. On the other hand, if the Baryon Asymmetry is generated by a 
high-scale mechanism of baryogenesis interesting consequences for the search of low energy lepton number violation and the
origin of neutrino masses can be deduced.}
\label{Con}
\end{figure}

By simply combining the arguments made above, we can conclude as follows:

If neutrinoless double beta decay is observed, it is:

\begin{itemize}
\item
{\it Either} due to a long-range mechanism, e.g. a light Majorana neutrino mass.

\item
{\it Or} due to a short-range mechanism. In this case it is very probable that lepton number is observed at the LHC.
This, however, implies that baryogenesis is a low-scale phenomenon which also may be observable at the LHC.
In this case there thus may well be a ``two-for-one'' deal at the LHC.
\end{itemize}

If, on the other hand, the BAU is generated at a high-scale, there will be no lepton number violation at the LHC.
If, in this case, neutrinoless double beta decay is observed, it thus will be typically due to a long-range mechanism. 
In combination with the assumption that we do not have a hint for lepton number violation at a low-scale in this case and on the
other hand a mechanism for the generation of the BAU at a high-scale, this will probably point towards a high-scale origin of
the neutrino mass as well, such as a vanilla-type seesaw mechanism in combination with leptogenesis.

Thus in summary an observation of neutrinoless double beta decay will typically (see Fig.~\ref{Con})

\begin{itemize}

\item
{\it Either} imply LNV at the LHC and low-scale baryogenesis and thus a possible observation of both processes in the near future.

\item
{\it Or} very probably a high-scale origin of both neutrino masses and baryogenesis.
\end{itemize}

We thus think that even if possible loopholes to these arguments may exist, it is important to 
stress these relations to make both model builders and experimentalists aware of the tight connections between 
neutrinoless double beta decay, the search for lepton number violation et the LHC and the origin of the Baryon Asymmetry of the Universe.

\bibliographystyle{ws-procs975x65}
\bibliography{ws-pro-sample}

%Non BiBTeX users can list down their references as:

\end{document}